\RequirePackage{fix-cm}
\documentclass[smallextended]{svjour3}       
\smartqed  
\usepackage{graphicx}
\usepackage{tabulary}
\usepackage{ tabularx}
\usepackage[utf8]{inputenc}
\begin{document}

\title{A Survey of Online Card Payment Fraud Detection using Data Mining-based Methods 
}


\author{Bemali Wickramanayake  \and 
        Dakshi Kapugama Geeganage \and Chun Ouyang \and Yue Xu 
}

\institute{B. Wickramanayake \at
              Business Intelligence, GVC Australia, Brisbane, Australia \\
              \email{bemali.wickramanayake@gmail.com}           
           \and
           D. Kapugama Geeganage \at
              Queensland University of Technology, Brisbane, Australia \\
							\email{dakshi.geeganage@hdr.qut.edu.au}
							\and
						C. Ouyang	\at
						Queensland University of Technology, Brisbane, Australia \\
							\email{c.ouyang@qut.edu.au}
							\and
							Y. Xu	\at
						Queensland University of Technology, Brisbane, Australia \\
							\email{yue.xu@qut.edu.au}
}

\date{}

\maketitle

\begin{abstract}
\vspace{-1.5\baselineskip}
Card payment fraud is a serious problem, and a roadblock for an optimally functioning digital economy, with cards (Debits and Credit) being the most popular digital payment method across the globe. 
Despite the occurrence of fraud could be relatively rare, the impact of fraud could be significant, especially on the cardholder. 
In the research, there have been many attempts to develop methods of detecting potentially fraudulent transactions based on data mining techniques, predominantly exploiting the developments in the space of machine learning over the last decade. 
This survey proposes a taxonomy based on existing research attempts and experiments, which mainly elaborates the approaches taken by researchers to incorporate the (i) business impact of fraud (and fraud detection) into their work , (ii) the feature engineering techniques that focus on cardholder behavioural profiling to separate fraudulent activities happening with the same card, and (iii) the adaptive efforts taken to address the changing nature of fraud. Further, there will be a comparative performance evaluation of classification algorithms used  and efforts of addressing class imbalance problem. Forty-five peer-reviewed papers published in the domain of card fraud detection between 2009 and 2020 were intensively reviewed to develop this paper. 
\end{abstract}

\section{Introduction}
\label{intro}
Online payment is a service consumed by many people and getting increasingly popular across the globe. Statistica \cite{1} is reporting that by the year 2020, the global digital payment penetration will be at 46\% with an expected average annual growth of 7.5\% over the next 4 years. Also, by 2020 the annual digital payment transaction volume is expected to be US \$ 4.4Tn and of that, ~50\% of the payments will be through credit and debit cards (directly) while ~44\% is through digital wallets that integrate credit and debit card and bank accounts. The popularity of online payments with credit and debit cards has also made it a lucrative area for fraudsters to make their gains, which is visible in global credit card fraud statistics. By 2019, the global fraud losses have accounted to US \$ 27.5Bn, according to PR Newswire Association LLC \cite{2} which is approximately 0.7\% of the digital transaction volume.

When analyzing the online card payment frauds, we can see 4 common types of them as per Mymoid \cite{3}. Friendly Frauds: This happens when a cardholder makes a digital purchase with their credit card, and then contacts their credit card issuer to dispute the charge. This can happen for genuine reasons (e.g.: booking a hotel online and not being happy with the condition once the cardholder arrives at the location), and also due to fraudulent reasons. Triangulation: This is when a fraudster sets up a fake merchant (generally a high bargain store or something that can lure an unsuspecting cardholder) and steals cardholder’s data once they make a purchase. Then the fraudster uses this data to make other online purchases. CNP (Card not Present) Fraud: the fraudster acquires real information about the cardholder (payment details and other individually recognizable details about the cardholder) from sources such as dark web, electronic scanning devices etc. This can happen when the payment facilitator and merchant both have compromised Data security, and hackers can easily obtain the data. In most cases, the fraudsters behave mimicking the buyer behaviour, such as spending time choosing and dropping a few items from the online cart, before buying a high-value item. Identity theft: here, the fraudster also obtains the personally identifiable data such as ID/ Passport info, age, likes/dislikes, family information, and behaves online the same way as the real customer does.

Detection of potential card payment fraud has become a topic of interest in payment gateways, payment aggregators, and banks to ensure optimum security for the end consumers and merchants who are dealing online. The popular payment gateways like Stripe \cite{4} are offering solutions to its merchants to insure the risk of card payment fraud by effective detection of fraudulent transactions in real-time. Fraud detection will not only save banks and payment gateways millions of dollars, but also could help to build better customer relationships. Indirectly, all sorts of payment fraud detection checkpoints adapted by Banks, Payment gateways, and E-Commerce platforms means, there will be a stronger and trustworthy digital economy for end consumers.

To address the card payment fraud detection there have been many researches and experiments that are conducted within academia and industry alike. Such methods include data mining-based techniques including machine learning and deep learning and external verification-based methods. In this paper, we evaluate the data mining-based methods and techniques that are proposed to solve the card payment fraud detection problem. 
When solving the card payment detection problem, it is notable that the researchers have mainly tried to address the following challenges/ problems.
\begin{enumerate}
	\item \textbf{Capturing Business Impact of Fraud:} Fraud detection in the card payment domain is an exercise with an inherent business impact. The direct business impact of “financial risk” for the cardholder and the bank/ merchant could be captured when this problem is framed as a cost-sensitive problem considering real-world costs. A problem is considered cost-sensitive when the cost of misclassification of a false positive is different from that of a false negative Ling and Sheng \cite{60}. Misclassification of a false positive generally consumes an administrative expense of investigation, while the cost of misclassifying a fraudulent transaction as non-fraudulent (false negative) can completely compromise the existing balance of a card.
	\item \textbf{Behavioural profiling to detect fraud patterns:} The behaviour of a fraudulent card transaction is rapidly adapting, making it difficult to tell apart a legal transaction and a fraudulent transaction. Seeja and Zareapoor \cite{46} highlight that it will be easier to profile what the legal transaction behaviour is, and then try to distinguish the fraudulent ones, than trying to find a common pattern among the fraudulent transactions.
\item \textbf{Class imbalance problem:} As highlighted by Japkowicz and Stephen \cite{64}, the class imbalance problem in classification refers to the extreme difference in the distribution of the prior classes within the dataset. Fraudulent transactions generally represent below  1\% of the total transaction population, which makes this is a highly imbalanced data set to tackle in a supervised learning problem. Out of the natural data sets used in the studies that are evaluated, the dataset used by de Sa et al. \cite{33} had the highest  percentage of fraudulent transactions of 1.84\% (obtained from a payment gateway), whilst the dataset used by E. Dunman and  Ozcelik \cite{21} had the lowest fraudulent transactions  percentage of 0.004\% (obtained from a Bank).
\end{enumerate}

With this survey, we expect to present a complete taxonomy and a comparative study of the solutions suggested by the research community in solving the card payment detection problem in different aspects as observed above. It is expected that a concise presentation of solutions proposed to the above challenges would help the future research work as well as the commercial application development efforts to systematically approach their work looking at the previous work. Further, a comparative performance evaluation of the algorithms that were used by them in their study is also presented.

Previous surveys done on similar subjects include the study conducted by Ryman-Tubb et al. \cite{52} which evaluates the card payment fraud detection techniques presented in the academia against the criteria that is critical in real-world applications. Abdallah et al. \cite{53} presented a study that provides a systematic overview of the challenges of fraud detection systems including credit card FDS such as concept drift, supports real-time detection, skewed distribution, a large amount of data. Ahmed et al. \cite{54} did a study on clustering-based anomaly detection techniques that are used in a variety of applications including financial fraud detection. Danenas \cite{55} discusses artificial intelligence techniques used in financial fraud detection including credit risk and payment fraud. Their paper's focus is more on a comparative study of the machine learning algorithms and techniques. Work presented by Sadgali et al. \cite{56} reviews the machine learning, deep learning, and other techniques used in financial fraud detection, focusing on the performance of methodology.

This survey will distinguish itself from those previous work, by addressing the key challenges of card payment fraud detection problem via a taxonomy which discusses how the researchers have tried to frame the business impact of the fraud detection, different approaches used to profile cardholder behaviour to distinguish fraud from the genuine transactions, techniques that have been used to combat the adaptive nature of the fraud. The analysis of classification mythologies used in these experiments will come as a secondary contribution of this paper. The proposed structure of this review is expected to help any future experiments in academia or in commercial application development, with its ability to provide a concise review of different approaches concerning the specific problems that are common in the payment fraud detection landscape. 

Hereinafter the paper will be organized as follows. In section 2, we briefly discuss our survey methodology and the extraction of sources. The main Taxonomy is introduced in Section 3. Sub-section 3.1 is dedicated for discussing the methods of addressing the business Impact of fraud and fraud detection. In sub-section 3.2 we discuss the feature selection methodologies using behavioural profiling. In 3.3 the adaptive techniques to handle the drift in fraud are discussed. A comparative study of the classifiers used in literature is presented in Section 4.  Then we briefly touch on the methods of handling class imbalance in section 5. The open areas that are unaddressed in the reviewed literature, or those which remain to be challenges are discussed in detail in section 6. After discussing the commercial applications in the domain, in section 7, we conclude the survey in section 8.
\section{Survey methodology}
\label{sec:1}
The objective of this survey is to provide an insight to the research community and commercial application developers who are interested in card payment detection problems on key challenge areas addressed via the contemporary research and experiments in machine learning and AI domain. Hence, A complete survey of the published work on or after 2009 related to machine learning and statistical methods for detecting credit card, payment card, or online payment frauds is included in this study. The relevant literature was explored in specific directories such as Science Direct and IEEE Explorer and Google Scholar was used to search for the literature might have been excluded in the directory specific searches. 

To select the papers, the following keywords were used. “credit card fraud detection”, “online payment fraud detection”, “e-commerce payment fraud detection”, “machine learning”, “AI”. When selecting the papers, the papers from the journals in Q1/Q2 and A*/ A conferences were given higher precedence. Figure \ref{fig:fig1} below depicts the distribution of papers, by year, and journal rank.

\begin{figure}
  \includegraphics[width=0.75\textwidth]{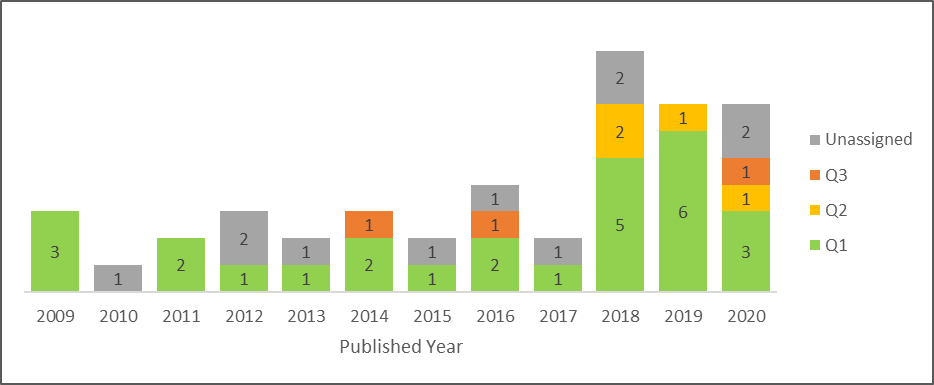}
\caption{Distribution of referred papers, by year, and journal rank}
\label{fig:fig1}       
\end{figure}

\section{Taxonomy and Review}
The motive to develop a taxonomy which is structured mainly based on the key challenges faced in terms of factoring the business impact, cardholder behaviour profiling based feature engineering was arose based on the practical challenges that had to be overcome during an effort to solve the payment card fraud for a real-world payment application.
Most of the previous studies focus mainly on the performance of the classification techniques, but we developed our taxonomy mostly around how a businessperson would approach the payment fraud detection problem. He would first need to identify how effectively does a solution help to minimize the business impact of fraud. In terms of business impact, we focus on how to factor the cost sensitivity into the fraud detection, and what are the initiatives taken to minimize the interruption to the business process by a fraud detection engine. Then, we investigate what are the alternative feature engineering approaches taken, to detect fraud by profiling the genuine cardholder behaviour. After the behavioural profiling, we investigate how one can handle the drift in fraud/ genuine transaction behaviour over the time. Outside the main taxonomy, we discuss the classification techniques, along with a comparative performance evaluation. Lastly, we briefly discuss the approaches taken to handle the class imbalance problem.

\begin{figure}
  \includegraphics[width=0.75\textwidth]{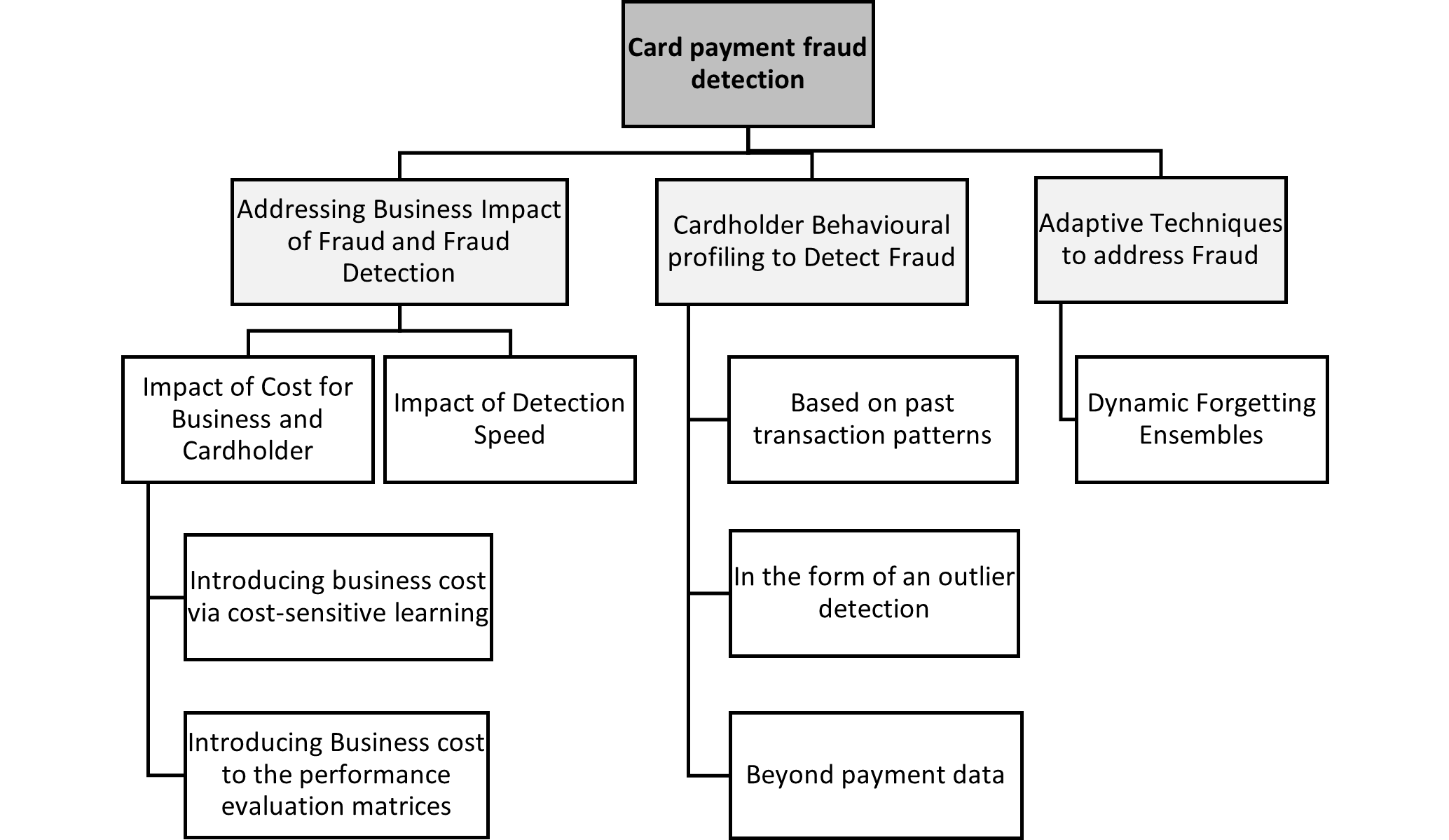}
\caption{Main Taxonomy}
\label{fig:fig2}       
\end{figure}

Each of the sections in taxonomy is elaborated with a thorough review of how the researchers have approached them with alternative solutions. Also, we bring in our perspective in the form of a discussion on the pros and cons of those approaches. Figure \ref{fig:fig2} depicts the main taxonomy developed.

\subsection{Addressing Business Impact of Fraud and Fraud Detection}
Payment fraud detection problem is inherently a high impact, yet low frequent problem. This increases the need for better performance accuracy, which is challenged by the class imbalance in the data. In general, fraudulent transaction volume accounts for below1\% of the total transaction population. This makes the detection extremely difficult, combined with the increasingly adaptive nature of the fraudsters to mimic legal transaction behaviours. However, one misclassified fraudulent transaction can forsake a cardholder his entire credit balance, or a serious loss to the merchant at the other end depending on how soon the cardholder figures out an unauthorized transaction and responds vs how flexible the chargeback policy of their bank.

On the other hand, in an effort of improving the model recall, the likely consequence is the increase of the false-positive rate. A high false-positive rate can also have a serious business impact, which involves the cost of investigation and deteriorated cardholder experience. Thus, an increase in false-positive rate means that the alerts generated by the Fraud Detection System (FDS) may not be used for effective real-time screening and actioning, without significantly impacting the user experience of the payment system.

The third area that is of high business impact is model latency. A real-time FDS is an extension to the standard payment processing path and adds a time cost to the same. This results in an additional delay to the time taken to process a cardholder’s transaction. Therefore, higher the latency of the FDS, higher the time taken to process a transaction, resulting in higher cardholder dissatisfaction. Although, may not be comparable 100\% like with like, it is noted by the online marketing community that, average page load time on Desktop is 5 seconds, and any further increase of the same results in serious bounce-back rates Mach metrics speed blog \cite{6}. This shows how important is it the ‘speed’ in the online activity domain, which is quite applicable for transaction processing as well. 

A typical anatomy of a real-time FDS is as follows.

\begin{figure}
  \includegraphics[width=0.75\textwidth]{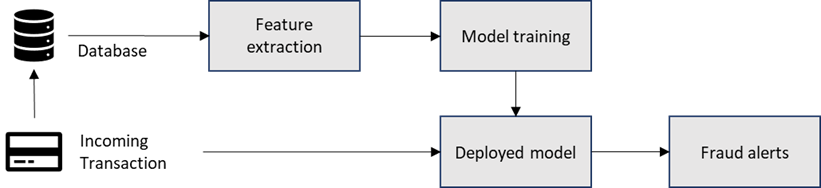}
\caption{Reat-time FDS}
\label{fig:fig3}       
\end{figure}

Considering the above, researchers have focused on improving fraud detection models by considering the cost sensitivity of the problem and latency as well.

\subsubsection{Impact of Cost for Business and Cardholder}

\begin{flushleft}
\textit{Introducing business cost via cost-sensitive learning}
\end{flushleft}

Cost-sensitive learning is an effort to introduce the impact of misclassification on the learning algorithm to minimize the overall cost. Shultz et al \cite{60} Cost-sensitive learning has been popular in detecting payment card fraud because the cost of misclassification of a fraudulent transaction and non-fraudulent transaction is significantly different.

When training a model, a cost function is introduced to the learning algorithm, which is a combination of the probability estimation, and the cost of misclassification. Then the model trains by trying to minimize the modified cost function, which also captures a penalty for real-world business cost. In the payment fraud detection problem, it is observed that the researchers have derived the cost function, by using different approaches to calculate the cost of misclassification. The cost of misclassification is the tradeoff between the likely loss of not detecting a fraudulent transaction, and the administrative cost of handling a legal transaction that is erroneously classified as fraudulent. When reviewing the approaches, we can see predominantly two ways of calculating the cost of misclassification by the researchers.

\begin{flushleft}
\textbf{Based on the instance value}
\end{flushleft}

Using instance value (transaction value) to calculate the cost of misclassification is demonstrated in the work by Olowookere and Adewale  \cite{7}, Bahnsen et al. \cite{13}, Noghani and Moattar \cite{28}, de Sa et al. \cite{33}, Kim et al. \cite{35} and Nami and  Shajari \cite{40}. The intuition behind choosing this parameter is that the impact of the misclassifying a fraudulent transaction is limited to that transaction itself. 

Using the transaction value-based cost matrix to identify the cost at a payment processing company/ e-commerce company level. Bahnsen et al. \cite{13} and Noghani and Moattar \cite{28} use a cost-sensitive learning in decision tree-based approach, based on instance level objective function based on cost savings, which is based on the transaction amount. In these methods, researchers have used cost function as a means of calculating the information gain of the features in the decision tree, to design the optimum tree structure. de Sa et al. \cite{33} presented a custom Bayesian network that was developed via a hyper-heuristic algorithm, that was cost-sensitive. In their work they have used the cost matrix in two different ways; instance re-weighing and adjusting the probability threshold, to minimize the overall misclassification cost. In these studies, the fraud detection is done at a payment processing company level, which justifies the selection of cost function. A payment processer or an e-commerce merchant is impacted by card payment fraud due to the chargebacks that follows a fraudulent transaction. Chargebacks result in additional processing effort, a fee, and a cash flow burden in certain cases CHARGEBACKS911 \cite{5}.

Transaction value has also been used in fraud detection in the context of a bank-level investigation. Kim et al. \cite{35} used cost-sensitive learning in a deep learning model, where the cost of misclassification is considered as the approved amount for false negative and the cost of contacting the cardholder for false positive. Nami and Shajari \cite{40} proposed cost-sensitive modelling has been done using Dynamic Random Forest with the Minimum Risk model, which is trained considering a loss function to represent the misclassification error. The cost of misclassification is considered as the transaction amount for false negative and the cost of action for false positive. Olowookere and Adewale \cite{7} presented a method of using a penalty for misclassification in an ensemble model, at the final classifier level (logistic regression) to introduce cost sensitivity at the algorithmic level. The cost function was derived from associating the “transaction value”. 

\textbf{Pros and cons: } As a clear advantage, we can say the use of transaction value as an algorithm level cost matrix will weigh the algorithm to detect those frauds that are significant in terms of the impact to the merchant, payment processor, and the bank. Also, it is an indicator of the cost that is most convenient to obtain, as it is a field that is available in almost all of the payment datasets. However, when the instance value (transaction value) is only considered, there is a high risk that early frauds through a stolen card may get unnoticed. Fraudsters may tend to test the stolen card information with one or two test transactions that help them to establish they are safe to carry out fraud with the same card for larger amounts. These transactions may not get flagged with a cost matrix that considers only the transaction value.

\begin{flushleft}
\textbf{Based on the available credit limit of the card}
\end{flushleft}

The intuition behind this method is that once a card is subject to a fraudulent transaction, it is an indicator that the entire balance / available credit limit of the card is compromised to fraud before the cardholder is aware of that fraud. Mahmoudi and Dunman \cite{12}, Sahin et al. \cite{15}, Dunman and  Ozcelik \cite{21}, and Duman and  Elikucuk \cite{50} exploited the advantage of this cost matrix to better weigh the impact of fraud from the cardholder and card issuer (bank)’s perspective.

Mahmoudi and Dunman \cite{12} first time used modified Fisher Discriminant Analysis, modified with a cost (weight) matrix based upon the available credit limit of the card. Researchers had worked with multiple cost matrixes using the available credit limit as the determining factor, to evaluate the best cost matrix.

\begin{equation}
\frac{Available usable limit (of the instance)}{Average available usable limit of all instnaces}
\end{equation}

\begin{equation}
{(1 + \frac{Available usable limit (of the instance)}{Average available usable limit of all instnaces})}^2
\end{equation}

\begin{equation}
{(1 + \frac{Available usable limit (of the instance)}{Average available usable limit of all instnaces})}^\frac{1}{2}
\end{equation}

\begin{equation}
{(ln (e + \frac{Available usable limit (of the instance)}{Average available usable limit of all instnaces})}^\frac{1}{2}
\end{equation}

\begin{equation}
e^{\frac{Available usable limit (of the instance)}{Average available usable limit of all instnaces})}
\end{equation}

And out of the above functions, cost function derived with equation 1 reported the best performance in terms of maximizing model recall and minimizing false positives. However, cost function derived with equation 2 had the best performance in terms of profit maximization when evaluated by the cost-saving of the top transactions. Sahin et al. \cite{15} proposed a transaction level cost matrix for the induction of the decision tree based on the available credit limit of the card to compute the cost of the false negative. Here the researchers are incorporating the cost in 4 distinct cost functions in the experiments. A) Direct cost Method, where the best fit is measured by reducing the overall cost when splitting the decision tree B)  Class probability method, where the relative frequency of the classes (class probabilities) are integrated in the cost calculation functions to add the effect of the class distributions to the node costs C) Gini method, in which the square of class probabilities are considered. D) Information gain method, in which negative logarithm of relative class frequencies are integrated in the cost calculation function. Out of the four methods, the last 3 methods have yielded superior performance in terms of model recall. Dunman and Ozcelik \cite{21} proposed a genetic algorithm-based optimization mythology to improve the fraud prediction model (heuristic/ rule-based), which used a cost matrix to train. As the cost of misclassifying a fraud as a legal transaction, they considered the available limit of the card. However, they also penalized the false positives with a higher cost of administration, which is double as the cost for a true positive.  Later Duman and Elikucuk \cite{50} also used the same cost matrix to train a model that was optimized using migrating bird’s optimization. 

\textbf{Pros and cons:} The obvious advantage is that this parameter considers the overall potential loss to the cardholder and card issuing entity (the bank) by the fraud. Despite the transaction value being small or large, the frauds that are susceptible to significant future impact is weighed higher with this technique. However, this may not be a piece of information that can be obtained when designing a fraud detection engine at a payment processor/ or an e-commerce merchant’s level. Also, in terms of the impact considered the detection algorithm may behave more favourable towards the cardholder and the bank, but not towards the payment processor/ e-commerce merchants whose main concern is the value of the transaction.

\begin{flushleft}
\textit{Introducing Business cost to the performance evaluation matrices}
\end{flushleft}

Card payment fraud detection is a binary classification problem. However, the results of the same have a significant commercial impact. Considering the same, the performance of the model is better evaluated through a commercial performance matrix, as opposed to the standard evaluation matrices such as model recall, AUC-ROC, or F1 measure.

Mahmoudi and Dunman \cite{12}, Bahnsen et al. \cite{13}, Sahin et al. \cite{15}, Gomez et al. \cite{18}, Dunman and  Ozcelik \cite{21}, Salazar et al. \cite{45}, Duman and  Elikucuk \cite{50} used cost-saving/ saved loss rate as a performance matrix to evaluate the model performance. Mahmoudi and Dunman \cite{12} used the percentage  of saved credit card limit (by detecting a fraud), considering the total balance available in the card.  Bahnsen et al. \cite{13} used the \% saved considering the transaction amount. Sahin et al. \cite{15} used the available credit limit to compute the saved loss rate. All three include the cost of handling a fraud alert into the computation, to include the real cost of false-positive alarms as well. Gomez et al. \cite{18} uses transaction amount as the cost of fraud. And considers the cost of false positives as a separate matrix. Dunman and Ozcelik \cite{21} and Duman and Elikucuk \cite{50} used the available credit limit, with a monitoring cost for false positives. Salazar et al. \cite{45} used Value Detection Rate (VDR) as a performance parameter which is calculated based on transaction value and measured False Positive Rate (FPR) separately. They had a target of achieving 0.5\% false positives, considering the transaction volume.

\textbf{Pros and cons:} The main advantage of using a cost-sensitive performance matrix is its ability to weigh the commercial impact of the decision that is provided by fraud detection engines. In the business world, the commercial impact is given a higher consideration when it comes to solution implementation. However, in terms of optimizing an algorithm, completely focusing on cost-sensitive performance matrices may overlook the ability to develop a superior version, which performs better in detecting most of the fraud instances irrespective of the cost.

\subsubsection{Impact of Detection Speed}
Pollazo et al. [9] observed that static learning would perform better in terms of speed, compared to dynamic learning models. Dheepa and Dhanapal [25] separated the fraud detection system into two sections, real-time vs non-real-time. The real-time model consisted of simple if-then type of rules, which takes less than ~200ms to process. The transaction authorization happens after this level of screening, and advanced feature-based fraud detection happens at a non-real-time model, which takes over ~1minute. Sherly \cite{47} Proposed using BOAT adaptive Decision tree classifier, by converting numerical features into categorical features using K-Means clustering, which is claimed to have superior performance in terms of speed compared to other decision tree algorithms. A standard decision tree algorithm operates by scanning the database for each level of the tree. However, BOAT algorithm can build several levels of the tree with a single scan of the database.

\textbf{Pros and cons:} As highlighted in the section introduction, low latency is an important feature in real-time fraud detection systems, to make sure better cardholder experience, especially in the online transaction domain. However, it should not compromise the model recall. Pollazo et al. [9] and Dheepa and Dhanapal [25] have highlighted that the speed was however achieved at the cost of a superior model performance. Thus, model latency could be considered as an open area of work, which will further be discussed under Open Issues.

\subsection{Cardholder Behavioural profiling to Detect Fraud}
\subsubsection{Behavioral profiling based on past transaction patterns}
Card payment fraud detection is an elaborate pattern recognition problem. The most frequent real-world scenario is that a fraudster would steal the card information from a legitimate cardholder and use that information to make a fraudulent online purchase of high value (Card not present fraud). Card not present fraud (CNP) amounts to nearly 90\% of the payment fraud in Australia as per the Australian Payment Network \cite{67}, where we could assume similar statistics prevail globally. Therefore, researchers have considered it as a more effective approach to figure out the past transaction behaviour of the legitimate cardholder, to isolate a fraudulent transaction.

This was achieved with the feature engineering techniques that consider past behaviour which considers the payment frequency, size of payments, the commodities purchased, etc.

\begin{flushleft}
\textit{Use of time aggregated features}
\end{flushleft}

Sanchez et al. \cite{8}, Jha et al. \cite{11}, Bahnsen et al. \cite{13}, Wu et al. \cite{19}, Askari and  Hussain \cite{23}, Zanin et al. \cite{24}, Dheepa and  Dhanapal \cite{25}, Jurgovsky et al. cite{29}, Lucas et al. \cite{32}, Carneiro et al. \cite{34}, Bhattacharyya et al. \cite{38}, Nami and  Shajari \cite{40} and Fu et al. \cite{44} have extensively used features that are aggregated over time axis to derive transactional behaviours of legitimate cardholders. A common observation that can be seen across the studies is, Transaction Amount and Count of Transactions that are aggregated over different periods, that was chosen by the researchers was a very popular way of gauging the past behaviour. It could have been the assumption that the amount that is spent by the genuine cardholders, as well as how many times they would use their card within a given period is expected to be fairly uniform for that particular cardholder, and any significant deviation would trigger an anomaly that could potentially be a fraud.

Following is a review of outstanding systematic feature engineering techniques that are used by some researchers in deriving time aggregated transaction behavioural features. Table 1 contains a summary of the time aggregated features found in the literature. Jha et al. \cite{11} and Bahnsen et al. \cite{13} have used aggregated parameters of value and count of transactions, across the time axis, subject to conditions, as a method of deriving historical payment behaviour patterns for each card. Jha et al. \cite{11} extensively used aggregated numerical features to explore the transaction history of both cardholder and merchant, across a multitude of periods (daily/ monthly/ over 3 months; quarterly). Bahnsen et al. \cite{13} proposed a systematic feature engineering framework to be used on top of raw transactional data, which included aggregated parameters, extended aggregated parameters (aggregated parameters that are subject to conditions such as a merchant, country or origination) and periodic parameters (which indicates whether a particular transaction falls within the confidence interval of a particular user's past transaction times) reflecting the transaction history. The aggregated features were calculated across an array of periods tp, where tp could be within the set of 1, 3, 6, 12, 18, 24, 72 and 168 hours. 

Jurgovsky et al. \cite{29} used Long Short- Term Memory (LSTM) Neural networks to identify the transaction sequences and automate the process of feature engineering, looking at the past transaction patterns. However, they also introduce a set of manually engineered features that represents the past transaction behaviour that includes (i) Number of transactions during past 24 hours, (ii) Number of transactions during past 24 hours, from the same merchant country, (iii) Number of transactions during past 24 hours, from the same merchant country, with the same merchant. Fu et al. \cite{44} used Convolutional neural networks (CNN) where they used the properties of visual recognition to derive patterns from a   large time aggregated feature matrix, where each element of the matrix represents the  “trading entropy” a particular transaction, which determines how different is the current transaction of the cardholder to the previous transactions (with any merchant) during previous period T. A large array of T’s was generated to use the feature matrix. This could be identified as one of the efforts where the researchers used CNNs creatively to select time aggregated features, which are often picked manually based on expert opinions.

Wu et al. \cite{19} utilize the FDA algorithm to derive features that considered the transaction data to be a time-dependent function of the card activity instead of a random discrete parameter, at the card level. The basic assumption behind this is that a card holder’s transaction pattern could be explained as a time-dependent function. The inclusion of FDA based features improved the results of the model significantly (recall from 0.954 to 0.970 and F1 score from 0.924 to 0.940)

\begin{table}
\caption{Number of time aggregated features used in each experiment (the number in the table refers to the number of features used in each study under each categorization)}
\label{tab:1}       
\resizebox{\textwidth}{!}{\begin{tabular}{|l|l|lllllllllll|}
\hline\noalign{\smallskip}
Raw Feature & Aggregation & \rotatebox{90}{Sanchez et al. [8]}  & \rotatebox{90}{Wu et al. [19]} & \rotatebox{90}{Askari and Hussain [23]} & \rotatebox{90}{Zanin et al. [24]} & \rotatebox{90}{Dheepa and Dhanapal [25]} & \rotatebox{90}{Jurgovsky et al. [29]} & \rotatebox{90}{Lucas et al. [32]} & \rotatebox{90}{Carneiro et al. [34]} & \rotatebox{90}{Bhattacharyya et al. [38]}& \rotatebox{90}{Nami and Shajari [40]} & \rotatebox{90}{Fu et al. [44]}\\
\noalign{\smallskip}\hline\noalign{\smallskip}
Country & Mode & & & & & & & & & & & n \\
Merchant\slash Store\_Info & Last & & & & 1 & & & & & & & \\
Merchant\slash Store\_Info & Count & 1 & & &  & & & & & & &  \\
Transaction\_Amount & Sum &  & 6&1 & & &1 &4 & &2 &3 &n  \\
Transaction\_Amount & Average &  &  &  & 1&1 & & & & 8& 3& n \\
Transaction\_Amount & Ratio & &5 & & & & & & & & & \\
Transaction\_Amount & Max & & 2& & & & & & & & &  \\
Transaction\_Amount & Standard deviation & & 1& & & & & & & & &  \\
Transaction\_Amount & Bias & & & & & & & & & & & n \\
Transaction\_Amount & Last & & & &1 & & & & & & &  \\
Transaction\_Amount & Previous & & & & & & & & & & 1& \\
Transaction\_Amount & Min & & 1& & & & & & & & &  \\
Transaction\_Count & Count &1 &7 & & & 1&1 &4 & &6 & 2& n \\
Transaction\_Amount & Ratio & & 8& & & & & & & & &  \\
Transaction\_Amount & Sum & & 1& & & & & & & & &  \\
Transaction\_Date\slash Time & Difference   & & &1 & 1& & 1& & 1& &1 &  \\
Transaction\_Date\slash Time & Average Difference & & & & 1 & & & & & & &  \\
Transaction\_processing & Mode & & & & & & & & & & & n \\
\hline
\end{tabular}}
\end{table}

\textbf{Pros and cons:} The clear advantage is time-dependent features, especially in terms of transaction amount and transaction count can be easily obtained to generate transaction patterns with regards to a cardholder. It could, for some extend uniquely define a cardholder behaviour, when evaluated across multiple sizes of periods generously (such as daily, weekly, monthly, every 3 months), since the spend is closely related to the personal income, which is more or less constant for most individuals during the short-medium term. However, there could be a significant loss of information when these parameters are obtained at the e-commerce merchant’s or a payment gateway’s level. This is because, the cardholders do not always use the same e-commerce merchant/ or payment gateway as a habit, unless it is used for mandatory periodic expenses such as bill/ subscription payment or grocery purchases. At Bank level, there could be a better visibility on a cardholder’s periodic spending pattern, however, is also susceptible to information loss in instances where the same cardholder has multiple cards with multiple banks, which is not a rare occurrence.

\begin{flushleft}
\textit{Payment Preference-based behavioural features}
\end{flushleft}

The payment behaviours that could represent the cardholder preference/ or the use case such as, to which merchant that he goes, what is the method of using the card, what kind of payments does he make using the card could give important clues that define a cardholder behaviour as well. Zhao et al. \cite{10}, Robinson and Aria \cite{14}, Zhang et al. \cite{22}, Zanin et al. \cite{24}, Nami and Shajari \cite{40}, Zheng et al. \cite{49} have defined feature engineering frameworks that consider such payment preference-based behavioural features.

Zhao et al. \cite{10} constructed transaction hierarchies combining multiple features that would define a transaction combining commodity => seller => buyer. In this approach, we can assume that the experiment was set-up to identify the most frequently occurring purchasing patterns of a particular buyer based on the implied “loyalty” or “familiarity” between a buyer and a seller, considering a particular commodity. 

\begin{figure}
  \includegraphics[width=0.75\textwidth]{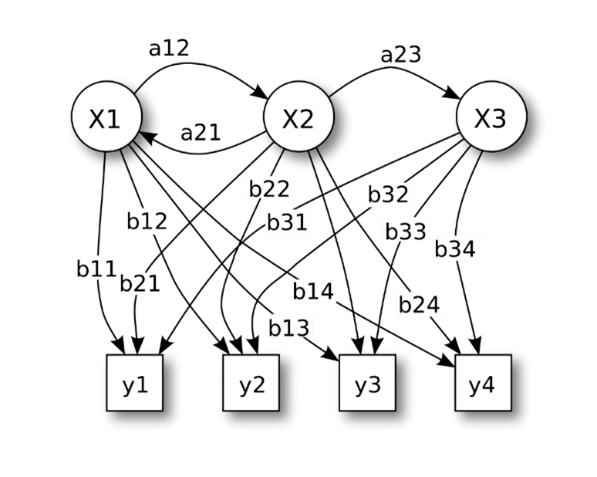}
\caption{Hidden Markov Model. Xi refers to the hidden states (what is not typically visible outside) and Yj refers to the observed states. ai represents the probabilistic relationships between hidden states, and bj represents the probabilistic relationships between hidden states and observed states. \cite{68}}
\label{fig:fig4}       
\end{figure}

Robinson and  Aria \cite{14} and Lucas et al. \cite{32} used Hidden Markov Model to evaluate the probability of a sequence of transactions, based on the features such as transaction type, merchant and transaction size. In the approach proposed by Robinson and Aria \cite{14}, HMM is trained with the past (legitimate transactions) and builds a probability matrix between its observed and hidden states.  in these methods, they have used the terminal used to perform the transaction (electronic/ grocery and miscellaneous) as the hidden state and the transaction size (cluster) high, medium, and low as the observed feature/state to construct the HMM. Once the probability matrix is trained, it is re-calculated with the new transactions being incorporated. And based on the difference of probabilities, it is decided whether the incoming transaction is a fraud or a legitimate transaction. This is an improved approach to the work presented by Srivastava et al. \cite{57}.  Robinson and Aria \cite{14} performed this at store level granularity, whereas Srivastava et al. \cite{57} performed this experiment at the card level granularity. Lucas et al. \cite{32} used HMMs for the sole purpose of automated feature engineering, and those features were later tested with an array of classification models. In their approach, they classified the incoming training transaction sequences into 3 types, where each type is represented by a binary representation. (i) transaction contains a fraud or not, (ii) transaction is obtained fixing the cardholder/ payment terminal, (iii) transaction is significant in terms of the amount of elapsed time compared to the last transaction in the sequence. Based on the combinations of the above 3 types, they trained 8 HMM models to obtain features resembling the likelihood of a particular transaction becoming a fraud. Figure \ref{fig:fig4} depicts the structure of HMM.

Zhang et al. \cite{22} proposed a novel method of feature engineering, compared to traditional aggregation-based feature engineering based on a defined Homogeneity Oriented Behavior Analysis (HOBA) in the transaction space. When developing the features, the researchers segregated the transactions into different kinds of transaction behaviours. Purchases, Refunds, Subscriptions, Cash withdrawals. The feature engineering had been done, isolating each space, at the card level. They proposed a systematic framework for developing aggregated features to represent the historical behaviour which is outlined by aggregation characteristic, aggregation period, transaction behaviour, and aggregation statistic (sum, mean, std deviation).
Instead of limiting to one or two manually selected features, the researchers used the automatic feature selection capabilities offered by the deep nets, with a large feature matrix as derived above. Zanin et al. \cite{24} proposed a feature engineering method based on parenclitic networks that not only evaluates the relationship between the features and the class (fraudulent/ not), but also intra feature relationships and how different/ same are they with that of fraudulent transactions. The motive behind this is to infer a classification model based on the differences that exist between fraudulent and legitimate transactions, based on the intra feature relationships. Nami and Shajari \cite{40} used the K-Nearest neighbours’ algorithm to derive a similarity feature between a cardholder's new transaction vs previous transactions. Zheng et al. \cite{49} proposed a method of user behaviour profiling based on 5 attributes (transaction time, transaction location, product, amount, shipping address) for each user, assigning transaction transition path probabilities to develop logical behavioural profile for each user.

\textbf{Pros and cons:} When evaluating the pros and cons of using user preference based behavioural features to establish the patterns, we first notice the following advantages. User preference, such as at which merchant he generally buys a particular product commodity, what is the average amount he spends at a particular merchant, how many subscriptions he has will help to define a user profile at a more robust level, than merely considering the transaction sizes and amounts. Naturally, users tend to establish a pattern based on the above parameters which are driven by the individual lifestyles. However, these parameters may not be accessible at all levels of the payment chain, and at different levels, there will be a loss of information. For an example, at bank-level, we will have the visibility of at which merchants do the cardholder make payments but not what products they buy, at e-commerce store level, although it is easier to establish what products does a cardholder buy, it lacks the visibility of at what other merchants does the cardholder make purchases. Another disadvantage is, such profiling may flag the genuine transactions that are made because of certain events of a cardholder’s life. Such as travelling abroad or a sudden sickness.

\vspace*{-10pt}

\subsubsection{In the form of an outlier detection}
We can also find several studies that treat payment fraud detection as an outlier detection problem. Outlier detection is a popular technique to detect fraud or defects, which may be used in conjunction with other behavioural features. Carcillo et al. \cite{36}, Olszewski \cite{39}, and Wang and Han \cite{42} have used outliers to differentiate fraudulent transactions from genuine transactions.
Olszewski \cite{39} used self-organizing maps (SOM) to visualize the user profiles and establish payment patterns of different user profiles at the account level, by converting the high dimensional feature set into a two-dimensional map based on the feature relationships. And then, the feature centroids have been calculated to represent the “most common behaviour” of an account. Fraudulent transactions were detected using a threshold based on the distance to the SOM centroid, treating the problem as an outlier detection problem. Carcillo et al. \cite{36} and Wang and Han \cite{42} used K-means clustering club transactional features and use the cluster means as a reference to derive transactional outliers that could help outstanding features that separate fraudulent transactions from nom fraudulent transactions in supervised learning models. Carcillo et al. \cite{36} experimented with 3 levels, global, local (card level), and cluster level (by combining multiple cardholders to a single cluster of similar behaviour). Out of the three levels, they observed that identifying outliers concerning a cluster that combines multiple cardholders performed the best, compared to the other 2 approaches. 

\textbf{Pros and cons:} Outlier detection is a statistical method, which is systematic. It is explainable to an end-user, in terms of the logic and the purpose it serves (which is identifying fraudulent behaviour, that is assumed to behave significantly different to the legitimate transactions, at least in a few aspects). However, common challenges of outlier detection as highlighted by Han et al. [58] are, how the normal behaviour is defined, and how to define the threshold to identify a particular transaction as an outlier.

\subsubsection{Beyond payment data}
Another intriguing yet effective effort in trying to model cardholder behaviour is considering the data that is beyond the scope of payments. Sanchez et al. \cite{8}, Zhao et al. \cite{10}, Lucas et al. \cite{32}, Carneiro et al. \cite{34} have used such parameters to profile a cardholder, and their behaviour to develop clues that could help in detecting fraud.

Sanchez et al. \cite{8} presented an approach, that is quite different from most of the other approaches seen in the payment fraud detection domain, distinguished by the fact that no payment transaction-related features are being considered. They develop an association rule pool based on the following features (i) product type (ii) place of purchase (iii) purchaser’s age (iv) purchaser’s sex (v) years account held (vi) no-money-down purchase (vii) purchase in maximum instalments (viii) purchase period. Notably, the demographic parameters including the cardholder’s age and gender have become prominent in their analysis, which indicates that they have concluded certain demographic groups can be more susceptible to fraud. With the exploratory data analyses they have observed that women and young people are more prone to fraud, the products that are purchased through a fraudulent transaction may easily be converted to cash, and some business types are more prone to fraud than the others as well. Further Lucas et al. \cite{32} and Carneiro et al. \cite{34} also have considered cardholder demography-based features in their work. Zhao et al. \cite{10} proposed a method of combining the implicit online behaviours and explicit transaction features to detect fraud, that is suitable for e-commerce platforms. The implicit features included the time taken by the buyer to complete a transaction, depth of pages that the buyer visits looking for a product and the time taken to browse alternative options, and how many previous visits has the buyer performed. The use of these features is justified by the ability of such features to model a legitimate buyer. For example, they state that a legitimate buyer tends to spend some time on the store before completing a transaction, and he/she will have visited the store many times before. Besides, we can also hypothesize that such features may have the possibility of detecting a fraudulent transaction that is made by a stolen card from a cardholder who has already visited this e-commerce platform, because the fraudster’s online buying behaviour will quite be different to the original cardholder.

Table \ref{tab:2}  below depicts the ‘beyond payment’ features that have been used in this context.

\begin{table}
\caption{Beyond Payment Features}
\label{tab:2}       
\resizebox{\textwidth}{!}{\begin{tabular}{|l|l|l|}
\hline\noalign{\smallskip}
References & Feature Used & Feature Category\\
\noalign{\smallskip}\hline\noalign{\smallskip}
Sanchez et al. [8]	&Cardholder’scurrentage	&Cardholder\_Demography\\
&	‘‘M”: male; ‘‘F”: female &	Cardholder\_Demography\\
Zhao et al. [10]	& Confirmation interval (time till buyer confirms delivery) & 	E-Commerce\_Online\_Behaviour \\
&	Rating interval of a transaction (how soon did the buyer rated)	& E-Commerce\_Online\_Behaviour \\
&	Average Staytime on the store prior transaction	& E-Commerce\_Online\_Behaviour \\
&	Average depth on the store prior transaction &	E-Commerce\_Online\_Behaviour \\
&	Real-time authentication of buyer	& E-Commerce\_Online\_Behaviour \\
&	Buyer rating &	E-Commerce\_Online\_Behaviour \\
&	Positive rating ratio given by the buyer &	E-Commerce\_Online\_Behaviour \\
&	Average confirmation interval of the buyer&	E-Commerce\_Online\_Behaviour\\
&	Average rating interval of the buyer&	E-Commerce\_Online\_Behaviour\\
&	Invalid rating ratio of the buyer	&E-Commerce\_Online\_Behaviour\\
Lucas et al. [32]	& its age	&Cardholder\_Demography \\
	& its gender	&Cardholder\_Demography \\
Carneiro et al. [34]& GenderMen	&Cardholder\_Demography \\
	&GenderWomen	&Cardholder\_Demography\\
	&GenderOther	&Cardholder\_Demography\\
\hline
\end{tabular}}
\end{table}

\textbf{Pros and cons:} Beyond payment data provide an additional layer of information for profiling the cardholders more effectively. The demographic information and associated the behavioural patterns displayed in online shopping improves the distinguishability between the genuine cardholder, and a fraudster who has the card information/ physical card of him/her, which helps to detect fraud. Also, such parameters may not be replicated/ imitated by the fraudster to mimic the original cardholder behaviour conveniently. The main challenge rather than a disadvantage is the availability of such data. Except for the instances where strict personal information disclosure requirements exist (e.g: when opening a bank account), the cardholder demography related parameters are not convenient to obtain. In situations wherein a household, one card is commonly used (genuine scenario), such data that is provided at the e-commerce terminal may not be consistent across. Also, accessibility of e-commerce behaviour related implicit data is often restricted to that e-commerce platform/ store only.

\subsection{Adaptive techniques to combat the changing nature of fraud}
Payment fraud is an evolving game. Fraudsters come across sophisticated methods to mimic the genuine cardholder behaviour in an attempt of not getting noted by the fraud detection systems. Thus, the fraud detection systems always in a battle to update themselves to detect new patterns of fraud. Beyond the standard approach of model retraining, we can see several attempts by Pollazo et al. \cite{9} and Seeja and  Zareapoor \cite{46} to overcome this challenge in the way their detection models were designed. 

\subsubsection{Dynamic Forgetting Ensemble}
Pollazo et al. \cite{9} used card level aggregated features (details are not disclosed) along with a “dynamic forgetting” ensemble where several classifiers are trained using semi overlapping chunks of time, the data set of the immediate next chunk of the time also contained the ‘predicted’ label based on the previous model that is developed. And the combined result is used to predict the fraud in the test data. When the test data time frame is moved ahead, the training data time frame chunks are also moved at the same length in time. The significance of this method is that whilst it adapts to the new fraud patterns, it also accounts for the fact of what past behaviours that need to be forgotten (whilst retaining the information on the behaviours that are still valid). This provides an incremental learning ability to the ensemble compared to the complete training data set replacement.

Seeja and  Zareapoor \cite{46} proposed a pattern mining technique where the model trains by retaining the most frequently occurring fraud and legal transaction patterns of each cardholder (frequent itemset mining) and classifying the incoming transactions using a matching algorithm against the retained patterns. The model is updated by updating the frequent patterns concerning each cardholder. With this approach, it has been made it challenging to the fraudsters to mimic the genuine cardholders, without knowing the exact cardholder’s buying behaviour.

\section{Comparative Study of Classifiers Based on Published Performances}
\begin{figure}
  \includegraphics[width=0.75\textwidth]{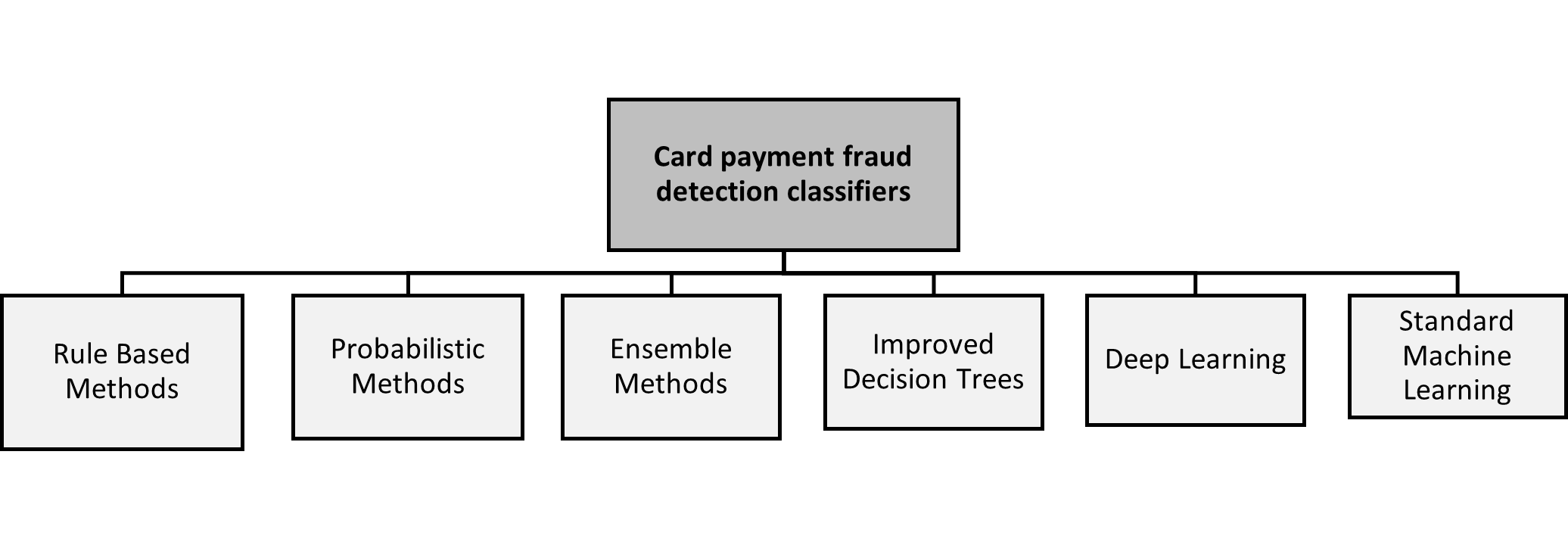}
\caption{Card payment fraud detection classifiers}
\label{fig:fig5}       
\end{figure}

In this section, we discuss different classification methods used by researchers in solving the credit card fraud detection problem. The performance of these models and algorithms is comparatively evaluated based on the published results. However, the limitation is that we cannot directly compare the performance of these methods, one with the other. Hence, we use the improvement of the method compared with the choice of baseline/ baselines used by the respective researchers as a way of performance comparison.

However, our focus will be on the novel methods proposed by the researchers, to discuss in detail. Towards the end of this section, we also acknowledge the standard classifiers that have been used in literature, along with a relative performance comparison.

\subsection{Rule-based methods}
Rule-based methods use a weighted matrix of rules and thresholds to determine if a transaction is fraudulent or genuine. The significance of rule-based methods is the superior model interpretability that is important in commercial-grade applications. Sanchez et al. \cite{8}, Dunman and Ozcelik \cite{21}, Gianini et al. \cite{26}, and Duman and Elikucuk \cite{50} have exploited rule-based methods in solving payment fraud detection problem.

Sanchez et al. \cite{8} developed a set of 50 fuzzy association rules to predict fraudulent transactions with over 67 percent confidence. Such rules were defined by combining multiple features that were discussed in section 3.2 in detail. The rules are linguistic, rather than numerical threshold-based, which makes it more intuitive for human experts. Dunman and Ozcelik \cite{21} used a genetic algorithm and scatter search to re-calibrate the weights used by an existing rule-based fraud detection system. Later the members from the same research group used migrating bird’s optimization to improve the same classifier further Duman and Elikucuk \cite{50}. Gianini et al. \cite{26} proposed a rule-pool based method, relying on the shapely value (which is the average marginal contribution from different rules to the final decision), which is a concept in game theory. Further, the researchers tried to minimize the rule pool to reduce computational complexity, by using three algorithms. (i) based on the individual rule's contribution to making the correct decision, based on past data. (ii) based on the shapely value itself, (iii) based on how the game (prediction) changes when we drop each rule. 

\textbf{Performance:} Table \ref{tab:3} depicts the comparative performance of rule-based techniques that were used.

\begin{table}
\caption{Rule-Based Techniques}
\label{tab:3}       
\resizebox{\textwidth}{!}{\begin{tabular}{|l|l|l|l|l|l|l|l|}
\hline\noalign{\smallskip}
Reference	&Method	&Performance &Performance	&Performance 	&Best &	Performance 	&Improvement  \\
& & Matrix	& &of the & Baseline & of the  & compared to  \\
& & & &baseline & & baseline & best baseline\\
& & & &(Best) & &  (Average) & \\
\hline\noalign{\smallskip}
Dunman and 	&Rule-based  	&Cost Saving	&0.9092	& & & & \\
Ozcelik [21] & - optimized & & & & & &  \\
& with GASS & & & & & &  \\
\hline\noalign{\smallskip}
Gianini et al. [26]	&Rule-based &	Recall	&0.3000	&0.3000	&Rules 	&0.3000	&0.0000\\
&- Shapely  & & & & combined & &\\
& value-based & & & & with OR & &\\
\hline\noalign{\smallskip}
Duman and 	& Rule-based 	&Recall&	\textbf{0.9194}&	0.7934&	Dunman and &	0.7934&	\textbf{0.1260}\\
Elikucuk [50] & - Optimized  & 	Cost Saving& \textbf{	0.9474}&	0.9092	&Ozcelik [21]&	0.9092	&\textbf{0.0382}\\
	& with MB& & & & & &\\
\hline
\end{tabular}}
\end{table}

\textbf{Pros and Cons:} The most significant advantage of rule-based methods is that they are easier to be interpreted by humans. Higher interpretability combined with a high level of model performance means that there will be better confidence in the results. The second advantage is that once the rule-based framework is developed, it is a less complex framework to implement in a live environment. However, the complex rule construction involved in the model development phase means, the effort involved in model updating will be high. If the model is not sufficiently updated to capture the most recent patterns in fraud (and genuine behaviours) the chances of the model being outdated are much higher.

\subsection{Probabilistic/ Statistical Methods}
Zhao et al. [10], de Sa et al. \cite{33}, Panigrahi et al. \cite{41}, and Zheng et al. \cite{49} used statistical theories-based models to detect card payment frauds. These methods largely exploit the conditional probability-based theories, which construct a probabilistic relationship between the features and class (fraud/ not fraud).Zhao et al. \cite{10} performed basic belief assignment to the features to construct evidence and used a modified approach to Dempster-Shafer theory to combine such evidences to infer a class label (fraudulent or not) to a transaction. When combining evidences, to optimize the weights assigned to each evidence, they used a basic genetic algorithm. de Sa et al. \cite{33} used a customized Bayesian network which is optimized using a hyper-heuristic evolutionary algorithm. Panigrahi et al. \cite{41} developed a four-component fraud detection system consisting with a rule-based filter, Dempster -Shafer adder to combine the rules and give reasoning with a probabilistic estimate P(h), Transaction history database to evaluate the duration from last transaction (D) and a Bayesian Learner to classify fraud with P(h|D) conditional probability. Zheng et al. \cite{49} used a probability-based logical Behavior method that was derived using 5 attributes (transaction time, transaction location, product, amount, shipping address) for each user,  assigning transaction transition path probabilities to develop logical behavioural profile for each user as discussed in section 3.2.2 above, and matching that with the incoming transaction to detect fraud - classify based on incoming transaction acceptance degree for a particular user profile.

\textbf{Performance:} Table \ref{tab:4} depicts the comparative performance of probabilistic learning techniques that were used.

\begin{table}
\caption{Performance of Probabilistic Learning Techniques}
\label{tab:4}       
\resizebox{\textwidth}{!}{\begin{tabular}{|l|l|l|l|l|l|l|l|}
\hline\noalign{\smallskip}
Reference	&Method	&Performance &Performance	&Performance 	&Best &	Performance 	&Improvement  \\
& & Matrix	& &of the & Baseline & of the  & compared to  \\
& & & &baseline & & baseline & best \\
& & & &(Best) & &  (Average) & baseline\\
\hline\noalign{\smallskip}
Zhao et al.[10]  & DS  	& Recall	& \textbf{0.8300}	& 0.5415 &	DS(e)	&0.2348	& \textbf{0.2885} \\
& Uncertainty & False Positive Rate & \textbf{0.0240} &	0.0331 &	Dong 	& 0.1267	& \textbf{0.0091}  \\
& Reasoning & & & &et al. [67] & &  \\
\hline\noalign{\smallskip}
de Sa et al.[33] 	&Fraud &	F1	&0.8270 &	0.8540&	SVM	&0.7384&	-0.0270\\
 &BNC & & & & & & \\
\hline
Panigrahi et & 	Bayesian &	Recall	& \textbf{+0.2000}	& & 	Card & &	 	\textbf{0.2000} \\
al. [41] & Learner	&	False Positive Rate	&\textbf{-0.0500}	& & Watch NN	& &	 	\textbf{0.0500}\\
\hline		
Zheng et al. [49]& 	logical & 	Recall	&\textbf{0.9500} &	0.8300	&HMM	&0.6667	&\textbf{0.1200}\\
 &	Behavioural & & & & & &  \\
& profile-based &	AUC-ROC	&\textbf{0.8400}	&0.7500	&HMM	&0.6600&	\textbf{0.0900}\\
 & method & & & & & &  \\
\hline
\end{tabular}}
\end{table}

\subsection{Ensemble approaches}
Ensemble learning involves combining multiple machine learning algorithms (called base learners in ensemble learning) to obtain a superior result than any of the base learners used. Such base learners are combined using techniques such as bagging, boosting, or majority voting to derive the result. The base learners can be either homogeneous or may belong to different classes of algorithms \cite{59}. In the work proposed in the card payment fraud detection area, we can observe that ensemble learning has been a popular choice. Pollazo et al. \cite{9}, Bagga et al. \cite{16}, Carta et al. \cite{20}, Zareapoor and Shamsoulmoali \cite{31}, Carcillo et al. \cite{36}, Wang and  Han \cite{42}, Sherly \cite{47}, Xuan et al. \cite{48}, and Kültür and  Çağlayan \cite{51} have used this approach.

Pollazo et al. \cite{9} experimented with multiple versions of dynamic ensembles based on 3 algorithms (Rand Forest (RF), Support Vector Machines (SVM), Neural Networks (NNET)) to feed the final decision using x models that were trained across the past x chunks of time as discussed in section 3.3. Bagga et al. \cite{16} proposed two methods for building an ensemble using 6 base classifiers, (i) to use pipelining to assemble feature selection algorithms with the classifier (Random Forest was used) and (ii) to use bagging classifier as a meta ensemble learner of which the base classifier will be random forest. Carta et al. \cite{20} also proposed an ensemble approach based on probabilistic threshold and majority voting, which combines the all or combinations of MLP, Gaussian Naïve Bias, Adaptive Boosting, Gradient Boosting and Random Forest models. Then ensemble models were proven to have better ROC-AUC, than the individual models. Zareapoor and Shamsoulmoali \cite{31} proposed a bagging ensemble decision tree to improve the performance of fraud detection, and the individual DT algorithm of the ensemble was J48 based on C4.5 type. Xuan et al. \cite{48} tested the performance of a CART based tree ensemble (which uses the Gini impurity of attributes to distribute data) against Random Forest to detect credit card fraud. Kültür and  Çağlayan \cite{51} proposed an ensemble combining 6 state of the art supervised algorithms, and evaluated the result using optimistic voting, pessimistic voting and weighted voting mechanisms. The algorithms used were Decision Trees (DT), Random Forest (RF), Bayesian network, Naïve Bias (NB), Support Vector Machines (SVM), and K-Nearest Neighbors (KNN).

Instead of using multiple supervised learning models to create ensembles, some researchers experimented with hybrid learning technique, with the focus of isolating the outlier behaviours in transactions. Carcillo et al. \cite{36} Used transactional outlier scores obtained from unsupervised learning methods to train a supervised learning algorithm, to combat the changing fraud behaviour. The outliers were derived at 3 levels of granularity. (i) Global, considering the entire transaction space, (ii) Local, at card level, (iii) Cluster granularity, by generating clusters of cards (cardholders) with similar transaction behaviours. It was concluded that the cluster granularity (with 10 defining clusters) performed to be the best approach in this experiment. Wang and Han \cite{42} combined supervised and unsupervised learning techniques via an integrated learning framework using K-Means clustering (unsupervised) and SVM/ Adaboost (supervised) algorithms. Sherly \cite{47} used hybrid learning, combining K-means clustering and Decision trees. However, the objective of using clustering techniques was to convert numerical features into categorical features that could be handled by the DT.

\textbf{Performance:} Table \ref{tab:5} depicts the relative performance of each of the ensemble learning approaches that are discussed above.

\begin{table}
\caption{Performance of Probabilistic Learning Techniques}
\label{tab:5}       
\resizebox{\textwidth}{!}{\begin{tabular}{|l|l|l|l|l|l|l|l|}
\hline\noalign{\smallskip}
Reference	&Method	&Performance &Performance	&Performance 	&Best &	Performance 	&Improvement  \\
& & Matrix	& &of the & Baseline & of the  & compared to  \\
& & & &baseline & & baseline & best \\
& & & &(Best) & &  (Average) & baseline\\
\hline\noalign{\smallskip}
Bagga et al. [16]	& Supervised & Recall	& 0.8700	& 0.9100	& LR,&  	0.8488	&-0.0400\\
& Random & & &  & Adaboost,& & \\
& Forest & & &  & QDA & & \\
& Ensemble &	& &  &  & & \\
\hline
Bagga et al. [16]	&Supervised &	F1 Score	&\textbf{0.8100}&	0.4600 &	Random &	0.2812 &	\textbf{0.3500} \\
& Random & & &  & Forest& & \\
& Forest & & &  &  & & \\
& Ensemble &	& &  &  & & \\
\hline
Carta et al. [20]	&Supervised & 	Recall&	0.8070&	0.9700	&Adaptive &	0.7992	&-0.1630	 \\
&	 	Ensemble  & & &  & Boosting& & \\
&(Best configuration)& & &  & & & \\ 
&- GBC, RFC & & &  & & & \\
\hline
Carta et al. [20]	&Supervised & 	AUC-ROC	&\textbf{0.8980}	&0.8960	&Random &	0.7586	& \textbf{0.0020}	 \\
&	 	Ensemble  & & &  & Forest& & \\
&(Best configuration)& & &  & & & \\ 
&- GBC, RFC & & &  & & & \\
\hline
Zareapoor and &	Supervised &  	Recall	& \textbf{0.8500}	& 0.5800& 	Naïve & 	0.5367	&\textbf{0.2700}\\
Shamsoulmoali [31] &Bagging & & &  & Bias & & \\
&Decision Tree & & &  &  & & \\ 
& Ensemble & & &  &  & & \\
\hline 
Xuan et al. [48]	&Supervised &	Accuracy &	0.9300	 & & & &  \\	 	 	 
& Random &  &	 &		 & & &  \\	
& Forest and & 		Recall&	0.6700 & & & &  \\	 	
&  Cart Tree & 		&	 & & & &  \\
&Ensemble	 & 		&	 & & & &  \\
\hline
Kültür and & 	Supervised & 	Recall	& \textbf{0.9392} & 	0.9257 & 	Naïve & 	0.6433 &	\textbf{0.0135}	 \\	 
Çağlayan [51] & Ensemble &  & & & Bias & & \\
\hline
Carcillo et al. [36] & 	Hybrid & 	AUC-PR	&\textbf{0.2400}&	0.2230&	Random &	0.2230	&\textbf{0.0170}	 \\	 
 & Ensemble &  & & & Forest & & \\
\hline
Wang and  Han [42]	&Hybrid &	AUC-ROC&	\textbf{0.7949}&	0.7755&	SVM&	0.7755&	\textbf{0.0194} \\
 & Ensemble &  & & &  & & \\
\hline
Sherly [47]	& Hybrid & 	Recall	& 0.8500 & & & &\\	 	 	 	 
	&Ensemble&	False Positive	&0.1000	& & & &\\	 
& & Rate &	& & & &\\
\hline
\end{tabular}}
\end{table}
\textbf{Pros and cons:} With ensemble learning, it is expected that it will generate superior results compared to any of the base learners, by combining the performance of each model. However, this could also be counterproductive when there are many weak learners combined with the strong learners. In such a scenario adopting an ensemble method such as majority voting may overpower the strong learners by the weak learner. We could see this in the results that were published in the study of Carta et al. \cite{20}, where the random forest-based model performed better than or equal to any of the other ensembles proposed.

\subsection{Improved Decision Tree-based approaches}
Askari and Hussain [23] and Noghani and Moattar [28] have attempted to improve the classic Decision Tree to solve this problem. Askari and Hussain [23] proposed an intuitionistic fuzzy logic-based decision tree, by combining the Intuitionistic fuzzy logic with the C4.5 decision tree to achieve better results in e-transactional fraud detection. This method considers the concepts of degree of membership $\mu$, degree of non-membership $\vartheta$, and the degree of indeterminacy $\pi$ proposed in fuzzy logic to determine the splitting decisions at the nodes. The degree of non-membership and degree of indeterminacy reduces the number of false alarms, producing a more operationally effective result. Nogami and Moattar [28] proposed DT based and extended wrapper-based feature selection, combined with a cost-sensitive decision forest. Features were ranked by to chi-squared, gain, and Relief values, against the labelled data. Each feature was evaluated using a decision tree. The features that improved the DT accuracy or did not reduce it were selected for the final model.

\textbf{Performance:} Performance of Decision Tree-based techniques in comparison with the baseline approaches are depicted in Table \ref{tab:6} below

\begin{table}
\caption{Performance of Modified Decision Tree-based Techniques}
\label{tab:6}       
\resizebox{\textwidth}{!}{\begin{tabular}{|l|l|l|l|l|l|l|l|}
\hline\noalign{\smallskip}
Reference	&Method	&Performance &Performance	&Performance 	&Best &	Performance 	&Improvement  \\
& & Matrix	& &of the & Baseline & of the  & compared to  \\
& & & &baseline & & baseline & best \\
& & & &(Best) & &  (Average) & baseline\\
\hline\noalign{\smallskip}
Askari and  & Intuitionistic  	& Accuracy	&\textbf{0.9800}&	0.9775&RF&	0.9700&	\textbf{0.0025} \\ 
Hussain [23] & fuzzy&  & &	 &	 	& 	&   \\
& logic-based  & False   &	\textbf{0.0080}&	0.0500	& SVM &	0.1250&	\textbf{0.0420}\\
& decision & Positive Rate& & & & &  \\ 
& tree & & & & & &  \\ 
\hline\noalign{\smallskip}
Noghani and &Decision &	Recall	&0.9980	&0.9980	&JD4	&0.9944	&0.0000\\
 Moattar [28] &Tree & & & & & & \\
&ensemble & F1	&\textbf{0.9980}	&0.9800	&JD4	&0.9764	&\textbf{0.0180}\\
\hline
\end{tabular}}
\end{table}

\subsection{Deep Learning}
The use of Deep Learning and Neural nets are remarkably effective in the automated feature selection and filtering abilities, as they recognize the patterns from raw features more efficiently compared to machine learning models Goodfellow et al. \cite{61}. Multiple layers of neuron performing complex calculations that are hidden/ or may not be easily interpreted gives the Deep learning models a bad rapport of being a black box. However, we can see that some researchers have creatively used the properties of popular Deep learning algorithms to detect card payment fraud. Pollazo et al. \cite{9}, Gomez et al. \cite{18}, Zanin et al. \cite{24}, Misra et al. \cite{27}, Jurgovsky et al. \cite{29}, Fiore et al. \cite{30}, Kim et al. \cite{35}, Zhu et al. \cite{37}, Fu et al. \cite{44} have used Deep learning models or Deep learning ensembles.
In the dynamic learning ensemble proposed by Pollazo et al. \cite{9}, Neural networks were used as one of the base learners in comparison with Support Vector Machines (SVM) and Random Forests (RF). However, in the experimental setup, it was observed that RF performed much better than the neural network.

Gomez et al. \cite{18}, Zanin et al. \cite{24} and Misra et al. \cite{27} have used Multi-Layer Perceptron (MLP) classifier in their work. Gomez et al. \cite{18} used two-staged neural net-based architecture. A first stage consisting of two levels of Artificial Neural Network (ANN) based filters which are trained to reduce data imbalance by filtering out genuine transactions and MLP classifier for the final fraud classification. Zanin et al. \cite{24} used MLP classifier, which was fed by both raw features, as well as topological features that represent the cardholder behaviour using Parenclitic networks (this approach is discussed in section 3.2 in detail). Misra et al. \cite{27} used MLP classifiers, together with autoencoder based feature selection methodology.

Jurgovsky et al.\cite{29} used LSTM based architecture, due to the inherent ability of (Long Short-Term Memory) LSTM networks to detect the sequential properties of incoming transactions that allow better payment pattern recognition. Fiore et al. \cite{30} used a Generative Adversarial Network to synthesize the fraudulent transactions (positive class) to improve classification performance. Kim et al. \cite{35} compared the performance of multiple topologies of deep learning models against the "champion" ensemble model (which was a combination of shallow neural network, logistic regression, and a decision tree) that was used in a real-world FDS system. Zhu et al. \cite{37} used an optimized weighted extreme learning machine, optimized with the dandelion (Genetic) algorithm to offer better results against for an imbalanced data set. And the methodology is then tested with credit card fraud detection applications.  Fu et al. \cite{44} used Convolutional Neural Networks (CNN) by converting the credit card data into an extensive temporal feature matrix and training the model to detect fraud.

\textbf{Performance:} Performance of Deep Learning techniques in comparison with the baseline approaches are as follows. 

\begin{table}
\caption{Performance of Deep Learning Technique}
\label{tab:7}       
\resizebox{\textwidth}{!}{\begin{tabular}{|l|l|l|l|l|l|l|l|}
\hline\noalign{\smallskip}
Reference	&Method	&Performance &Performance	&Performance 	&Best &	Performance 	&Improvement  \\
& & Matrix	& &of the & Baseline & of the  & compared to  \\
& & & &baseline & & baseline & best \\
& & & &(Best) & &  (Average) & baseline\\
\hline\noalign{\smallskip}
Gomez et al. [18]	&MLP	&AUC-ROC&	0.8700		& & & & \\		
\hline
Zanin et al. [24]	&MLP&	AUC-ROC&	0.8300& & & & \\		
\hline
Gomez et al. [18]	&MLP&	Saved& 	0.2300 & & & & \\	
& & Loss Rate 	& & & & & \\
\hline
Misra et al. [27]&	MLP&	Recall	&\textbf{0.8015}&	0.7279	&KNN&	0.6324&	\textbf{0.0736}\\
\hline
Jurgovsky et al. [29]&	LSTM	&AUC-PR&	\textbf{0.2440*}&	0.2265*	&RF	&0.2265*&	\textbf{0.0175}\\
\hline
Fiore et al. [30]	&3-layer 	&Recall&	0.7303& & & & \\	
& ANN	& &	& & & & \\
\hline
Fiore et al. [30]	&3-layer 	&F1	&0.8106& & & & \\	
& ANN	& &	& & & & \\
\hline
Kim et al. [35]&	Deep NN	&F1	&\textbf{0.9275}&	0.9090&	Ensemble&		&\textbf{0.0185} \\
\hline
Zhu et al. [37]&	Optimized   &	AUC-ROC	&\textbf{0.8000}&	0.5900&	Unoptimized& 	&	\textbf{0.2100} \\
&WELM & & & & WELM & &\\
\hline
Fu et al. [44]&	CNN&	F1	&0.2900&	0.2900	&RF&	0.2800&	0.0000 \\
\hline
\end{tabular}}
\end{table}

\textbf{Pros and Cons:} The advantages of Deep Learning that were exploited by the researchers in this domain were, the ability to filter features Gomez et al. \cite{18}, ability to handle large feature matrices Fu et al.\cite{44} and intrinsically handling sequential properties of the transaction data that needs to be considered when detecting fraud Jurgovsky et al. \cite{29}. The main disadvantage is the low interpretability. Also, the neural networks with a large number of hidden layers introduce an additional computational burden that can result in a higher model latency.

\subsection{Standard Machine Learning Methods}
It is commonly observed that there are a vast number of experiments (of which the main contribution is towards feature engineering), exploit popular machine learning classification algorithms to perform the classification task. The most popular classifier that has been used is Random Forest (RF), and in terms of the performance also, we observe that RF is a better choice. Other algorithms that have been used frequently are Support Vector Machines, Decision Tree, Logistic Regression, XG Boost and AdaBoost.

\textbf{Performance:} Table \ref{tab:8}   below depicts the comparative performance of the studies where standard classifiers of Random Forest, Support Vector Machines, Decision Trees, Logistic Regression have been used with little or no modifications, as a part/ baseline or the main classifier of the experiment.

\begin{table}
\caption{Performance of Standard Classifiers}
\label{tab:8}       
\resizebox{\textwidth}{!}{\begin{tabular}{|l|l|l|l|l|l|l|l|}
\hline\noalign{\smallskip}
Reference & Performance & Random & Support & Decision & Logisitic & XG Boost & Adaboost \\
& Matrix & Forest & Vector&   Tree & Regression & &\\
& & & machines& & & &\\
\hline
Jurgovsky et al. [29]	& AUC-PR	& 0.2265* & & & & &\\
\hline
Carta et al. [20]	& AUC-ROC	& 0.8960 & & & & & \\
\hline
Zhang et al. [22]	& AUC-ROC	& \textbf{0.9490}	& 0.9410 & & & & \\
\hline
Carneiro et al. [34]	& AUC-ROC	& \textbf{0.9950}	& 0.9005	& &	0.9010 & & \\
\hline
Wang and Han [42]	& AUC-ROC	& & 0.7949 & &	& &	\textbf{0.9842} \\
\hline
Whitrow et al. [43]	& Custom Loss 	& \textbf{0.0550}	& 0.0600	& &	0.0570	& & \\
& Function** & 	& 	& &		& & \\
\hline
de Sa et al. [33]	& F1	& 0.7840	& 0\textbf{.8540} & &		0.7050	& &\\
\hline
Fu et al. [44] &	F1	& \textbf{0.2900} &	0.2700	& & & &\\
\hline
Jha et al. [11]	& False Positive	& & & &		0.0500	& &\\
\hline
Kültür and Çağlayan [51] & False Positive	& \textbf{0.0191}	& 0.0445 &	0.0265 & & &\\
\hline
Jha et al. [11]	& Recall	& & & &			0.8300	& &\\
\hline
Sahin et al. [15]	& Recall & & &			0.9280	& & &\\
\hline
Wu et al. [19]	& Recall	& \textbf{0.7730}	& 0.7530	& & &		0.7610 &\\
\hline
Carta et al. [20] &	Recall &	0.8070 & & & & &\\
\hline
Zhang et al. [22]	& Recall	& \textbf{0.4277} &	0.3240	& & & &\\
\hline
Zareapoor and & & & & & & & \\
Shamsoulmoali [31]	& Recall	& &	0.4800 & & & &\\
\hline
Wang and Han [42]	& Recall & &		0.5920	& & & &	\textbf{0.8370}\\
\hline
Seeja and Zareapoor [46]	& Recall &	\textbf{0.6250} & 	0.1000 & & & &\\
\hline
Kültür and Çağlayan [51] &	Recall	& 0.5084	& \textbf{0.6689}	& 0.5253	& & &\\
\hline
Bahnsen et al. [13]	& Saved Loss Rate &	\textbf{0.5000}	& &	&	0.1000 & & \\
\hline
Sahin et al. [15] &	Saved Loss Rate	& & &	0.9600	& & &\\	
\hline
\end{tabular}}
\end{table}

\section{Class Imbalance}
As highlighted in previous sections, card payment fraud is an infrequent occurrence of high impact. This results in significant class imbalance in any of the natural datasets that are obtained. When there is class imbalance prevailing in the data, machine learning algorithms are ineffective in recognizing patterns from the minority class. Treating the class imbalance at the training stage helps the model to get trained better with the fraudulent transaction patterns more effectively by remembering the distinct nature of the fraud. 

\subsection{Handling class imbalance by under-sampling the negative class}
In this method, researchers try to improve the ratio between fraudulent and legal transactions by under-sampling the negative (legal transactions) class. Whilst this helps to reduce class imbalance, this technique is likely to prevent the model from obtaining sufficient information about the legal transaction patterns by causing information loss, and likely to blur the decision boundary. 

Pollazo et al. \cite{9}, Sahin et al. \cite{15}, Dunman and  Ozcelik \cite{21}, Misra et al. \cite{27}, Zareapoor and Shamsoulmoali \cite{31}, Kim et al. \cite{35}, Bhattacharyya et al. \cite{38}, Wang and  Han \cite{42}, Whitrow et al. \cite{43}, Xuan et al. \cite{48} have used this technique to handle class imbalance. Pollazo et al. \cite{9} used stratified sampling to undersample the negative class, to reduce the ratio between the positive and negative classes from 1:22,500 to 1: 10.  Dunman and Ozcelik \cite{21} experimented with 3 levels of imbalance to improve the speed of training the genetic algorithm. They observed that high levels of imbalance meant that their algorithm took weeks to train and optimize.
Bhattacharyya et al. \cite{38} Further tried to evaluate the comparative performance of this technique by experimenting with training samples with multiple proportions of Fraud data percentage and observed that higher the percentage of fraud data, higher the model performance becomes. Xuan et al. \cite{48} also tested the impact of the proportion of fraud/ legal transactions in the training set for the performance. The optimum ratio to be found was 5:1.

\subsection{Oversampling the positive class by synthesizing fraudulent transactions}
Using techniques such as SMOTE to oversample the positive class is another method that researchers use, whilst preserving the information about the legal transactions as it is. With this method, it is expected to generate more samples which closely resembles the nature of minority class. However, to perform a synthesizing technique, it is required that all the features of the data is numerical. Hence it may not be a technique that could be applied in methods that heavily relies on categorical features. Olowookere and Adewale \cite{7}, Pollazo et al. \cite{9}, Salazar et al. \cite{45} have used SMOTE in their work, whilst Fu et al. \cite{44} have used a novel technique to synthesize the positive class closer to the decision boundary. However, there is also evidence that oversampling techniques can introduce overfitting problems Chawla et al. \cite{62}.

\section{Open Issues}
In the card payment fraud detection domain, few areas remain challenges for developing a perfect solution, that could safeguard the cardholders and institutions against fraud. Data collection, due to availability of data due to confidentiality and sufficiency of information. Data Labelling, due to the reliability of human labelling and Model Latency, due to its impact on commercial grade operation are the issues that are discussed in this section.

\subsection{Data Collection}
Data collection, in real-world applications, is not a very challenging problem for internal application development. However, as payment data is considered highly confidential, it is not an easy task to obtain a sufficiently rich data set, outside the organizational boundaries for research and development purposes. Also, what level of information is available within the collected data to solve the problem depends on the point of collecting data. The breadth of information that is available at the Bank level, Payment gateway/ payment processor level, and e-tail level is significantly different.  Especially when trying to collect data below the bank level, the unavailability of the full transaction history of the cardholder imposes serious limitations in developing a behavioural profile for the cardholder, when trying to distinguish fraud.

\subsection{Data labelling}
Fraud labelling happens as a manual exercise. And most of the time, the trigger to label a transaction as a fraud is a chargeback request from the cardholder. However, there is generally a latency (of nearly one month since the transaction) that is associated with originating a chargeback request from cardholder end, as well as there can be many instances where a card fraud may go unnoticed to the cardholder. Also, there can be instances where a chargeback request is raised, not due to payment fraud as well. Originally the chargeback concept was developed in e-commerce space to protect the end consumer from committing to a transaction for a good/ service that does not conform to the expected quality CHARGEBACKS911 \cite{5}. Thus, there can be occasions where non-fraudulent transactions are also being flagged as fraudulent based on the charge-back trigger. This results in an inherent challenge of training models precisely based on the transaction-based patterns, due to inaccurate class labelling.

\subsection{Model latency (inverse: Detection Speed)}
There have been some efforts of improving model latency that is discussed under section 3.1.2. However, in general, this is an area that is often overlooked, but extremely important in commercial-grade applications. The feature and model complexity results in high latency in fraud detection systems, which results in a long time to process and authenticate a transaction, when used in a serial configuration. Also, the additional dependencies on the relational database management systems that are used for transaction processing means a higher load on top of the existing process, as well as additional system vulnerabilities. 

\section{Commercial applications in the area}
Since payment fraud detection is an area with a high business impact, there are many popular commercial applications provided in the market. The applications that are developed internally to safeguard the operations and cardholders by large payment gateways and banks, also the applications that can be purchased and customized to suit the smaller operations such as payment aggregators and e-commerce merchants.  Here, with this section, the most prominent solutions adopted by the market leaders of the domain are discussed in brief.

\subsection{Offered by the payment gateways}
Stripe \cite{4}, and Paypal \cite{65} the two leading global payment gateways use machine learning-based solutions to protect the users and merchants who use their payment gateway from possible fraud. As one of the direct benefits that are claimed by both parties in managing the impact of the chargeback, which is cost-intensive both in terms of effort and cash. The details of their approaches however are not available.

\subsection{Customizable frameworks provided by cloud service providers}
Amazon Web Services (AWS) offers a customizable fraud detection framework, that is pre-configured to make commercial fraud detection application development efforts simpler Amazon Web Services, Inc \cite{66}. They offer an ability to develop a fully automated machine learning engine where the developer can completely focus on feature selection and algorithm selection, which is specific to the use case, even without a full understanding of the model deployment technologies. The final model could then be invoked via an AWS API.

\section{Conclusion}
Card Payment fraud can be considered as one of the biggest  barriers towards a worry-free digital economy. Whilst the volume of fraud is relatively smaller in comparison to the volume of online transactions that are happening, the individual impact of a card fraud could be significant, both to the cardholder and the merchant/ Bank. With this survey, we were able to identify the methodologies proposed by the research community to address the business impact of card payment fraud detection domain. The main areas considered were handling the cost sensitivity of the problem, as well as handling the speed of processing and transaction authentication. Further, we evaluated different approaches taken to profile the cardholder behaviour to enable machine learning models to distinguish fraudulent transactions better. We classified these methods based on the logic used to profile the cardholder, as well as the breadth of information used. We also discuss some of the efforts to handle the fraud drift (adaptive techniques). Then we moved into a comparative study of the classification algorithms used by the research community, and then a brief explanation of class imbalance handling techniques. With this approach, we expected to provide a complete view of alternative solutions/ approaches available for payment fraud detection system development, for the researchers in academia and commercial application developers, in relevance to the key problems that were identified.


%
%



\end{document}